\documentclass[10pt,letterpaper]{article}
\usepackage{opex3}
\usepackage{amsmath}
\usepackage{ae}
\usepackage{cite}

\begin{document}

%=================================================================================================
%%%%%%%%%%%%%%%%%%%%%%%%%%%%%%%%%%% Title %%%%%%%%%%%%%%%%%%%%%%%%%%%%%%%%%%%%%%%%%%%%%%%%%%%%%%%%
%=================================================================================================

\title{A high-speed multi-protocol quantum key distribution transmitter based on a dual-drive modulator}
\author{Boris~Korzh, Nino~Walenta, Raphael~Houlmann, and Hugo~Zbinden}
\address{{GAP-Optique, University of Geneva, CH-1211 Geneva 4, Switzerland}}
\email{boris.korzh@unige.ch}

%=================================================================================================
%%%%%%%%%%%%%%%%%%%%%%%%%%%%%%%%%%%% Abstract %%%%%%%%%%%%%%%%%%%%%%%%%%%%%%%%%%%%%%%%%%%%%%%%%%%%
%=================================================================================================
\begin{abstract}
We propose a novel source based on a dual-drive modulator that is adaptable and allows Alice to choose between various practical quantum key distribution (QKD) protocols depending on what receiver she is communicating with. Experimental results show that the proposed transmitter is suitable for implementation of the Bennett and Brassard 1984 (BB84), coherent one-way (COW) and differential phase shift (DPS) protocols with stable and low quantum bit error rate. This could become a useful component in network QKD, where multi-protocol capability is highly desirable.
\end{abstract}


%==================================================================================================
%%%%%%%%%%%%%%%%%%%%%%%%%%%%%%%%%%%% References %%%%%%%%%%%%%%%%%%%%%%%%%%%%%%%%%%%%%%%%%%%%%%%%%%%
%==================================================================================================

%====================================================================================
%%%%%%%%%%%%%%%%%%%%%%%%%%%%%%%%%% Introduction %%%%%%%%%%%%%%%%%%%%%%%%%%%%%%%%%%%%%%====================================================================================

\section{Introduction}
Quantum key distribution enables the distribution of provably secure shared bit strings, which is an important fundamental primitive for many cryptographic tasks such as one-time pad encrypted secure communication \cite{Shannon48} or message authentication \cite{Portmann2012}. In contrast to classical secret key distribution schemes, only quantum key distribution (QKD) \cite{Gisin2002b, Scarani09} has been proven to provide universally composable security against an arbitrarily powerful eavesdropper, who is only restricted by the laws of quantum physics \cite{BenOr05, Renner2005}. Since the first theoretical conception \cite{Wiesner1983,  Bennett84}, a wide variety of different QKD protocols have emerged and have been demonstrated in numerous real-world scenarios \cite{SECOQC, Chapuran09, Madrid09, Chen:10, SwissQuantum, Sasaki11}.

For implementations based on optical fiber telecommunication infrastructures, many QKD systems rely on quantum states prepared in the time-phase domain to benefit from their inherent robustness against polarization fluctuations during propagation through the fiber. Moreover, the receiver of such systems can be rendered completely passive, without the need for active components which would introduce losses in the quantum channel and require additional resources from a fast random-number generator. Among the most prominent of such QKD protocols are the BB84 \cite{Bennett84} time-phase coding scheme \cite{TomitaDDM07, Yoshino12}, the COW protocol \cite{stucki-2005-87, Branciard2008} and the DPS protocol \cite{DPS02, Liu:13}.

Many interesting approaches of establishing QKD between different users in a network scenario have been proposed and demonstrated. Mainly, they have adopted either the trusted node implementation \cite{SECOQC, SwissQuantum, Sasaki11}, passive optical switching \cite{Madrid09} or active optical switching \cite{Chapuran09, Chen:10}. The latter two methods allow the quantum channels to be dynamically reconfigured in order to connect different users without relying on trusted nodes. In all of the demonstrations so far, the physical layer of the QKD systems has been restricted to specialized transmitters and receivers running the same QKD protocol. However, in reconfigurable network environments there is a potential interest in the capability of switching between several QKD protocols. This can stem from a variety of reasons, such as certain protocols being better suited for long distance links, varying resistance to environmental effects or background noise, individual user demands or different industrial standards. These things are already evident from current network implementations \cite{Sasaki11}, where a broad range of commercial and development systems are being used. Having a single transmitter capable of performing QKD with receivers dedicated to different protocols could provide a reduction in system complexity, management and cost. In this work we aimed at developing such a transmitter suitable for deployment in a reconfigurable network as represented in Fig.~\ref{fig:DDMnetwork}.

In \cite{Tomita201055} it was shown that an arbitrary time-bin qubit state can be generated using a pulsed laser source, an unbalanced Mach-Zehnder interferometer (MZI) and a dual-drive modulator (DDM). In this work we apply and extend this result to implement a simplified DDM based QKD transmitter which does not require a stabilized interferometer. Moreover its absence allows for flexible time-bin period adjustment achieved by changing the DDM clock frequency. Such flexibility is essential when switching between receivers which may have a different interferometer path difference. This transmitter is especially suited for compact implementations of distributed-phase reference protocols, namely COW and DPS, as well as BB84 schemes based on weak coherent states with nonrandom phases \cite{LoPreskill07}. We present experimental results obtained with a GHz clocked multi-protocol QKD platform, which is based on a DDM transmitter and a reconfigurable receiver with free-running avalanche photodiode (APD) single photon detectors. A hardware-based key distillation engine is used to analyze the performance of the source in various QKD scenarios. Our transmitter exhibits stable low quantum bit error rates (QBER) in both the time and phase bases, suitable for multi-protocol QKD in optical fiber infrastructures. 

In the following, we will describe the operation of the DDM and how it can be used to prepare states required for the previously mentioned protocols. Note that it could also be applied to related protocols such as SARG \cite{SARG04} and B92 \cite{B92}, however we will not analyze them in this work. Experimental analysis of the state preparation is carried out, followed by incorporation of the transmitter in a QKD scenario. Overall performance is presented whilst using different protocols with various channel losses. The stability of the source is also analyzed over a long period of time. 

%========================================================================================
%%%%%%%%%%%%%%%%%%%%%%%%%%% State Preparation %%%%%%%%%%%%%%%%%%%%%%%%%%%%%%%%%%%%%%%%%%%
%========================================================================================

\section{QKD state preparation based on dual-drive modulation}\label{sec:state_prep}

%%%%%%%%%%%%%%%%%%%%%%%%%%% Fig. Network %%%%%%%%%%%%%%%%%%%%%%%%%%%%%%%%%%%%%%%%%%	
\begin{figure}[tbp]
\centering\includegraphics[width=10cm]{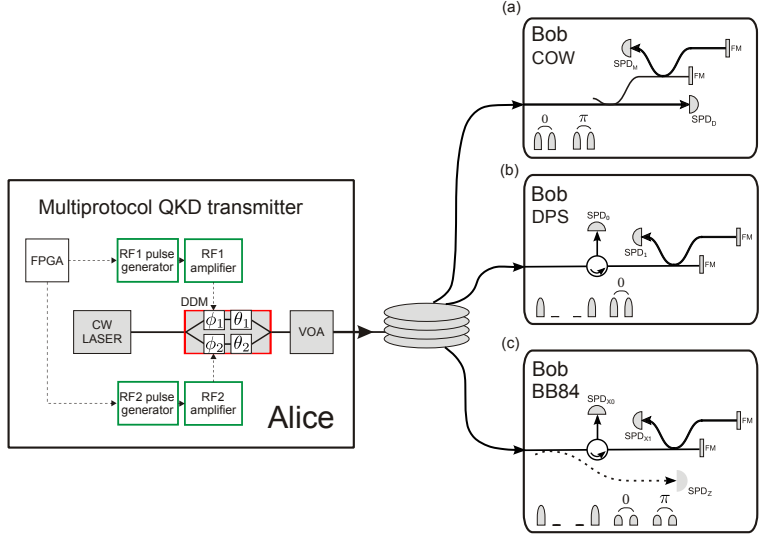}
\caption{Sketch of the multi-protocol QKD transmitter and of three potential QKD receivers running the (a) COW, (b) DPS or (c) BB84 protocols. The transmitter is composed of a dual-drive modulator which transforms the output of a continuous wave (CW) distributed feedback (DFB) diode laser, followed by a variable optical attenuator (VOA).  A field-programmable gate array (FPGA) drives two radio frequency (RF) pulse generators by pulse sequences required for the relevant protocols. The receiver is implemented using an interferometer with an optical path difference coresponding to a time delay $\tau$ and single photon detectors (SPD). The abstract component in between the transmitter and receivers represents an actively or passively switched network establishing the connection.}
\label{fig:DDMnetwork}
\end{figure}
%==================================================================================

A DDM consists of two electro-optic phase modulators, one in each arm of an integrated MZI, as illustrated in Fig.~\ref{fig:DDMnetwork}. The ability to control the phase shifts in the individual DDM arms enables the generation of arbitrary qubit states, limited only by the capability of the driving electronics. Each of the phase modulators provides an electrical input port to apply a DC bias voltage $V_{\text{DC},i}$ which leads to a fixed phase $\theta_i=\pi\cdot V_{\text{DC},i}/V_{\pi}$. Here, $V_{\pi}$ is the characteristic voltage which shifts the phase by $\pi$ and $i$ indicates the DDM arm. A second pair of radio frequency (RF) input ports allows the injection of voltage pulses of amplitudes $V_{\text{RF},i}$ which apply additional phase shifts $\phi_i=\pi\cdot V_{\text{RF},i}/V_{\pi}$. Throughout this paper it is assumed that the DDM has a Z-cut structure \cite{wooten2000}, hence a positive voltage induces a positive phase shift in both arms. Quantum models for electro-optic phase and amplitude modulation have been described in detail in \cite{Capmany:10pha, Capmany:10amp}. Accordingly, any coherent state $\left|\alpha\right\rangle$ with complex amplitude $\alpha$ at the input of the DDM is transformed as

\begin{equation} \label{eq:DDM_full}
\left|\alpha\right\rangle \rightarrow\frac{e^{i\left(\theta_{1}+\phi_{1}\right)}+e^{i\left(\theta_{2}+\phi_{2}\right)}}{2}\left|\alpha\right\rangle. 
\end{equation}
For simplicity, we choose the DC bias voltages such that $\theta_{1}=0$, $\theta_{2}=\pi$. Therefore the resulting state $\left|\psi\right\rangle$ is given by

\begin{equation} \label{eq:DDM_RF}
\left|\psi\right\rangle = \frac{e^{i\phi_{1}}-e^{i\phi_{2}}}{2}\left|\alpha\right\rangle, 
\end{equation}
hence for  $\phi_{1}=\phi_{2}$ the interference is destructive and no light passes through the modulator. Applying temporary RF voltage pulses with $\left|\phi_{1}-\phi_{2}\right|=\pi$ results in constructive interference and output pulses of maximum intensity are obtained. The global phase is determined by an appropriate choice of $\phi_{1}$ and $\phi_{2}$. If the input is a coherent continuous wave (CW) laser, it is possible to cut out multiple coherent pulses with an arbitrary phase relation, i.e. an arbitrary time-phase qubit (or qutrit or any higher dimensional state) can be prepared. For example, in order to generate two pulses with a phase difference of $\pi$, we first apply $\phi_{1}=0$ and $\phi_{2}=\pi$ and then, after a time $\tau$, $\phi_{1}=\pi$ and $\phi_{2}=0$. In order to operate the DDM transmitter at high frequencies it is desirable to utilise a method for generating the required qubit states by using RF drive signals with the minimum number of voltage levels, preferably just two.

In reality it is not possible to apply the desired phase shift instantaneously, since the RF drive pulses have a finite rise time and this leads us to consider the resulting phase chirp of pulses carved out from a CW laser. It is possible to obtain chirp-free intensity modulation by operating the DDM in the so called push-pull mode, where opposite phase shifts of equal magnitude are applied to the DDM arms simultaneously. This method can be used for production of the time and phase basis qubits, however if both bases are used, as for BB84, then 4 voltage levels would be required for driving the DDM arms. By allowing the pulses to be chirped, the number of voltage levels can be reduced, simplifying the drive electronics. Generally, increasing the number of voltage levels in high-speed electronics bears the drawback of increased signal eye spreading (amplitude fluctuations) \cite{Ho2005}. Such amplitude fluctuations could have a detrimental effect on the QBER. In the following we will show that having a phase chirp should not actually increase the QBER, meaning the overall system performance can be improved by using fewer voltage levels.

\begin{figure}[tbp]

\begin{minipage}[b]{0.5\textwidth}
		\centering
		\includegraphics[width=6.8cm]{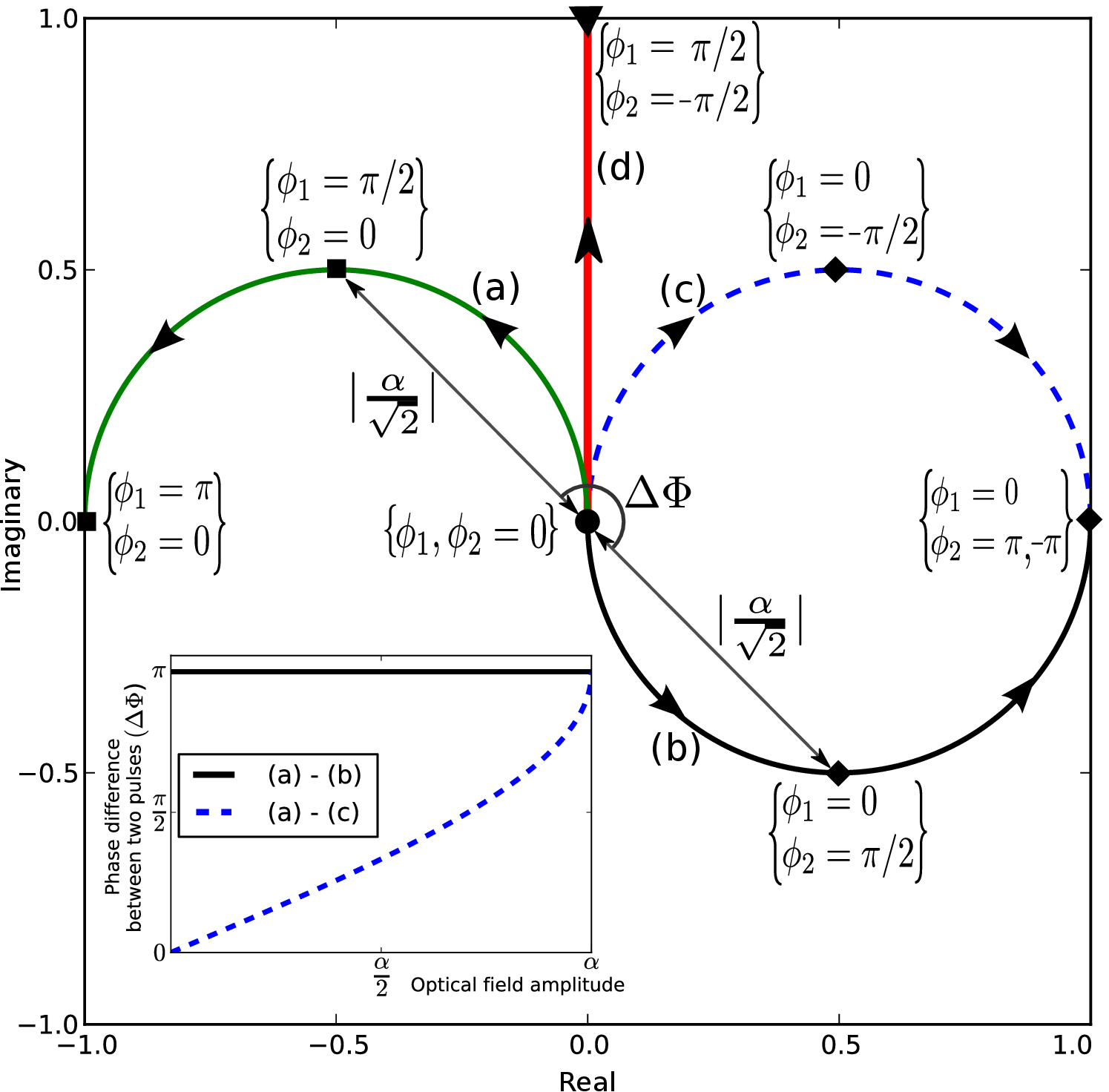}
	\end{minipage}
	\hspace{0.04cm}
	\begin{minipage}[b]{0.5\textwidth}
		\centering
		\includegraphics[width=6.8cm]{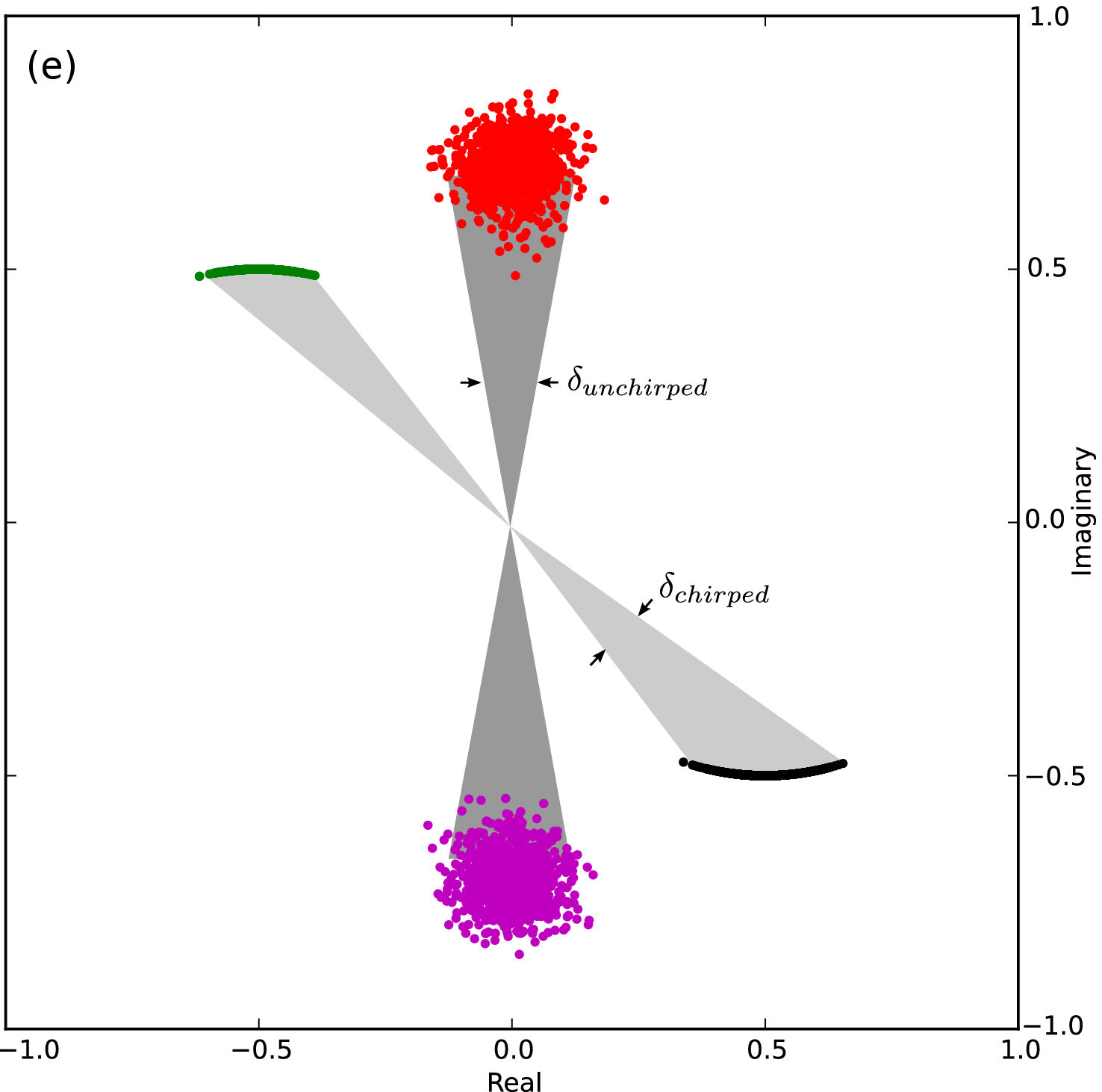}
	\end{minipage}

\caption{Complex plane representation for the rising edge progression of optical pulses generated by either (a) maintaining the phase shift of DDM arm 2 at zero, whilst applying a phase shift of $+\pi$ to arm 1 or (b) applying the $+\pi$ phase shift to arm 2 whilst keeping arm 1 at zero. Curve (c) shows the effect of applying a negative phase shift $-\pi$ in arm 1. (d) Illustrates push-pull operation achieved by applying $\phi_{1}=\pi/2$ and $\phi_{2}=-\pi/2$ simultaneously. The instantaneous field amplitude of the optical pulses is represented by the radial distance from the origin (with a maximum of $\alpha$), whilst the optical phase difference between two pulses is represented by the angle $\Delta \Phi$. \emph{Inset:} Optical phase difference as a function of optical field amplitude, between pulses (a) and (b) along with (a) and (c). The latter results in a varying phase difference, unsuitable for phase coding. (e) Illustrates a comparison of phase coding using the chirp free (push-pull) and chirped methods when operating with noisy multi-level drive signals.}
\label{fig:Constellation}

\end{figure}

Considering that the phase qubit is formed by two optical pulses separated by a time $\tau$, it is sufficient to ensure that each pair of points on the two pulses separated by $\tau$ have the correct relative phase, in other words the phase chirp is matched. In order to visualise the phase progression of the optical pulses, Eq.~(\ref{eq:DDM_RF}) is expanded into real and imaginary parts;

\begin{equation} \label{eq:DDM_real}
Re[\psi] = \frac{1}{2}\left[ \cos{\phi_{1}} - \cos{\phi_{2}}  \right]
\end{equation}
\begin{equation} \label{eq:DDM_imag}
Im[\psi] = \frac{1}{2}\left[ \sin{\phi_{1}} - \sin{\phi_{2}}  \right].
\end{equation}
Plotting Eqs.~(\ref{eq:DDM_real}) and (\ref{eq:DDM_imag}) on a complex-plane, Fig.~\ref{fig:Constellation} shows the rising edge progression of the optical field amplitude generated by applying an RF signal with an arbitrary rise time to either arm 1 or arm 2 of the DDM, i.e. varying $\phi_{1}$ or $\phi_{2}$ independently. Visualized like this, the optical phase between two pulses $\Delta \Phi$ is the relative angle between them and the optical field amplitude is the radial distance from the origin. It is clear that the pulse progression for a positive phase shift in arm 1 (trace (a) in Fig.~\ref{fig:Constellation}) is rotated by 180 degrees from that of a positive phase shift in arm 2 (trace (b) in Fig.~\ref{fig:Constellation}). Inset of Fig.~\ref{fig:Constellation} shows that the optical phase difference $\Delta \Phi$ between these two optical pulses is constant as a function of optical field amplitude. This means that two pulses generated by equal phase shifts (same rise-time and amplitude of RF signals) applied to different DDM arms would interfere destructively in an optimal way. Similarly, a pair or pulses originating from a phase shift in the same DDM arm (twice trace (a) in Fig.~\ref{fig:Constellation}) would interfere constructively since the pulses are identical. This enables the generation of phase encoded $\sigma_{\text{X}}$ qubits (phase difference of $\pi$ or 0 between two optical pulses) of arbitrary amplitudes. We wish to stress that this method of phase basis state generation has significant advantages over conventional methods for phase coding with a phase modulator, for example as used in \cite{Liu:13}, where the finite rise time of the phase modulator drive signal introduces errors and necessitates further temporal filtering. Our transmitter is inherently robust against this problem and reduces the number of components.

It should be noted that attempting to generate the $\sigma_{\text{Y}}$ states (phase difference $\pi/2$ or $-\pi/2$ between two optical pulses) with two level drive signals in this manner would lead to unmatched phase chirp and hence imperfect interference. This is illustrated with pulse progressions (a) and (c) in Fig.~\ref{fig:Constellation}, where a phase shift of $\phi_{1}=\pi/2$ is applied for the first pulse and $\phi_{2}=-\pi/2$ for the second. Inset of Fig.~\ref{fig:Constellation} shows the optical phase difference between these progressions, as a function of optical field amplitude, showing that the relative phase is not constant.

Pulse generation using the push-pull method is also illustrated by progression (d) in Fig.~\ref{fig:Constellation}, where phase shifts of $\phi_{1}=\pi/2$ and $\phi_{2}=-\pi/2$ are applied simultaneously. This is used for $\sigma_{\text{Z}}$  basis encoding by sending a pulse in either the early or late time-bin. In fact, inverting the applied phase shifts ($\phi_{1}=-\pi/2$ and $\phi_{2}=\pi/2$) would generate a pulse with a relative optical phase difference of $\pi$, hence as mentioned previously, it is possible to achieve phase encoding using the push-pull method. It is clear however that this requires an additional voltage level. In fact, switching between the phase and time basis, as in BB84, would require four voltage levels in order to change the pulse amplitude. One can expect that going from a two-level signal to a three-level signal can increase the RF signal eye spreading by a factor of 1.5, whilst using a four-level signal could increase it by a factor of 3 \cite{Ho2005}. In order to estimate the impact this would have on the QBER, let us consider drive signals that have an eye spreading of 10\% for two voltage levels. Figure~\ref{fig:Constellation}(e) shows a comparison of the two phase encoding schemes (assuming BB84) modeled with noisy drive signals. For the chirped method, one arm is driven with a two-level signal, whilst the second one uses a three-level signal (required for basis switching). In this scenario, the chirped method would result in around 2.8\% QBER in the phase basis, whist in the chirp free case it would be about 4.9\% due to the additional eye spreading of the four-level signals. This shows the benefit of opting for a coding scheme with the minimum number of voltage levels.  

%%%%%%%%%%%%%%%%%%%%%%%%%%% Fig. Constellation Diagram %%%%%%%%%%%%%%%%%%%%%%%%%%%%%%%%%%
%%%%%%%%%%%%%%%%%%================================================%%%%%%%%%%%%%%%%%%%%%%%

A simple and robust coding scheme for generating the required states for the BB84, COW and DPS protocols is illustrated in Fig.~\ref{fig:Coding}, showing the necessary RF drive signals in the two arms of the DDM and the resulting optical output. The status of the security proofs is very different for each protocol and it is out of the scope of this investigation to compare the relative advantages in detail, however we will highlight the specific security framework to be used for analysis of the achieved secret key rates in this work. In the following we give an overview of the three protocols:

%%%%%%%%%%%%%%%%%%%%%%%%%%% Fig. Coding %%%%%%%%%%%%%%%%%%%%%%%%%%%%%%%%%%%%%%%%%%%
\begin{figure}[tbp]

\centering\includegraphics[width=\textwidth]{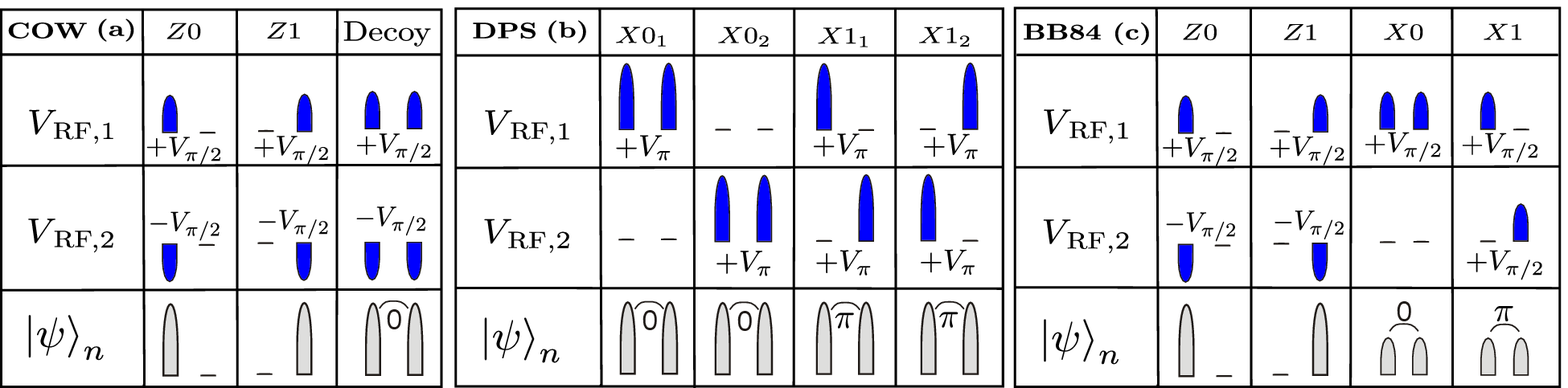}
\caption{Coding scheme for time-phase quantum state preparation, enabling the use of COW (a), DPS (b) and BB84 (c) protocols; Positive and negative electrical pulses (push-pull operation) in the two arms of the DDM produce the time basis states, whilst two consecutive positive pulses, either in the same arm or alternate arms, produce the phase basis states. All of the states for COW and DPS can be generated using DDM drive signals with only 2 voltage levels. BB84 requires a 3 level signal for arm 2 of the modulator, since the time and phase bases require negative and positive pulses in this arm, respectively.}
\label{fig:Coding}

\end{figure}
%==================================================================================

%%%%%%%%%%%%%%%%%%%%%%%%%%%%%%%% Protocol Descriptions %%%%%%%%%%%%%%%%%%%%%%%%%%%%%

\begin{description}

\item[COW]~(Figs.~\ref{fig:DDMnetwork}(a) and \ref{fig:Coding}(a)) encodes the bit values with a pulse emitted in the early or late time bin. This means in order to generate a bit value 0 or 1 at time $t$, we send the state  $\left|0\right\rangle_{t}\left|\alpha\right\rangle_{t-\tau}$ or $\left|\alpha\right\rangle_{t}\left|0\right\rangle_{t-\tau}$, respectively. For security reasons, a so called decoy sequence with two succeeding pulses is randomly introduced from time to time, which is represented by $\left|\alpha\right\rangle_{t}\left|\alpha\right\rangle_{t-\tau}$. Here, $\tau$ represents the imbalance of the interferometer on the receiver side. Note that in COW the relative phase between pulses is always the same, i.e. phase setting $\left|-\alpha\right\rangle$ is never used. The photon number per pulse is always constant, unlike BB84. Thus it is possible to generate the pulses either using the push-pull method utilizing both arms of the DDM, or by using only one arm and increasing the drive voltage to $V_{\pi}$. Bob determines the bit value by simply measuring the arrival times of the photons with a detector, know as the data single photon detector  ($\text{SPD}_{\text{D}}$). A beam splitter sends a fraction of the pulses to an unbalanced interferometer where the coherence of succeeding pulses is checked with detector $\text{SPD}_{\text{M}}$, which guarantees the security of the protocol. Similar to DPS, Bob can also perform measurements across bit separations to restrict the eavesdropper's attack strategies. This scheme is robust against the PNS attack and a security proof against some collective attacks has been given in \cite{Branciard2008}. In this work we will also take finite key effects into consideration for this protocol alone, as outlined in \cite{Tomamichel12}. 

\item[DPS]~(Figs.~\ref{fig:DDMnetwork}(b) and \ref{fig:Coding}(b)) uses a phase difference of 0 between two consecutive optical pulses to encode the bit value 0 and a phase difference of $\pi$ for the bit value 1. In order to generate a bit 0, we apply a double RF pulse in the same arm of the DDM to generate either $\left|\alpha\right\rangle_{t}\left|\alpha\right\rangle_{t-\tau}$ or $\left|-\alpha\right\rangle_{t}\left|-\alpha\right\rangle_{t-\tau}$. In order to generate a bit 1, we apply RF signals in opposite DDM arms to achieve $\left|-\alpha\right\rangle_{t}\left|\alpha\right\rangle_{t-\tau}$ or $\left|\alpha\right\rangle_{t}\left|-\alpha\right\rangle_{t-\tau}$. Bob has an unbalanced interferometer with two outputs adjusted in a way to obtain a click in detector $\text{SPD}_{\text{0}}$ or $\text{SPD}_{\text{1}}$ depending on the bit value. Note that in DPS we can encode two bits per time-bin pair since Bob can additionally measure the phase across qubits, i.e. the phase difference between $\left|\psi\right\rangle_{t-\tau}$ and $\left|\psi\right\rangle_{t-2\tau}$ which is always 0 or $\pi$ and leads to conclusive detections in $\text{SPD}_{\text{0}}$ or $\text{SPD}_{\text{1}}$. For calculation of the secret key rate we will use a security proof which is limited to individual attacks, outlined in \cite{Waks2006}.
	
\item[BB84]~(Figs. \ref{fig:DDMnetwork}(c) and \ref{fig:Coding}(c)) uses 4 states from 2 non-orthogonal bases. In the time-phase version, $\sigma_{\text{Z}}$ is the computational basis, having a pulse in the early or late time bin. These states are created as for COW to obtain $\left|0\right\rangle_{t}\left|\alpha\right\rangle_{t-\tau}$ and $\left|\alpha\right\rangle_{t}\left|0\right\rangle_{t-\tau}$. The states of $\sigma_{\text{X}}$ are $\left|\frac{\alpha}{\sqrt{2}}\right\rangle_{t}\left|\frac{\alpha}{\sqrt{2}}\right\rangle_{t-\tau}$ and $\left|\frac{-\alpha}{\sqrt{2}}\right\rangle_{t}\left|\frac{\alpha}{\sqrt{2}}\right\rangle_{t-\tau}$, where the pulses have to have half amplitude in order to maintain the same photon number per qubit. This scheme is similar to COW, where only one state of basis $\sigma_{\text{X}}$ is used for checking the security. BB84 uses both states and normally both bases with equal probabilities. Bob can use a measurement based on three detectors, where two are used to measure after the interferometer in the $\sigma_{\text{X}}$ basis, and the third performs a time-of-arrival measurement in the $\sigma_{\text{Z}}$ basis. Although not used in this work, an alternative detection scheme with only two detectors (labeled $\text{SPD}_{\text{X0}}$ and $\text{SPD}_{\text{X1}}$) can be used if at least one time bin is left empty between the qubits. Then, detection in the first or third time bin of either detector indicates the bit value after a measurement in $\sigma_{\text{Z}}$, whilst depending on the detector in which a photon was detected for the second time bin gives the bit value of a $\sigma_{\text{X}}$ measurement. This detection scheme was first proposed in \cite{Brendel1999} and is explained in further detail in \cite{Walton2005}. The security of the protocol is not altered since only the qubit frequency is changed. Note that in previous experiments \cite{ TomitaDDM07, Yoshino12, Marand:95} the BB84 protocol has been implemented with a pulsed laser source, an unbalanced interferometer identical to the one at Bob, plus a phase and/or intensity modulator. The advantage of these implementations is that the states are phase randomized.  It has been shown in \cite{LoPreskill07} that even using nonrandom phases, as in our case, still yields a secret key in place of general attacks, albeit at a much restricted distance. The secret key rate will be calculated using this security proof. The advantage of using our source, is that Alice can not only save the interferometer, but she can also generate almost any time delay $\tau$ with the electronic control system, making the transmitter easily adaptable to different receivers. In principle it could be possible to achieve phase randomization by pulsing the laser, bringing it below threshold between qubits, an approach that has already been demonstrated for high-speed random number generation \cite{Jofre11}. Otherwise, dual-drive modulators exist with an integrated phase modulator at the output, which could be used to actively phase randomize the qubits. As we have shown, the DDM is capable of producing phase and time basis states with arbitrary amplitude, hence in the future it would be possible to directly implement the decoy-state BB84 protocol \cite{wang2005}, by using multi-level RF drive signals. 
	
\end{description}
%================================================================================

%==================================================================================
%%%%%%%%%%%%%%%%%%%%%%%%%%% Experimental Demonstration %%%%%%%%%%%%%%%%%%%%%%%%%%%%
%==================================================================================
\section{Experimental demonstration}

Figure~\ref{fig:DDMnetwork} shows a sketch of the experimental setup for the source used to produce the quantum states from a continuous laser beam. We use a distributed feedback (DFB) diode laser (Anritsu) with a center wavelength of 1555~nm and a spectral line width below 3~MHz, hence a coherence time much longer than the typical delay $\tau$ between the time bins, which is usually on the order of nanoseconds. After the pulses are generated by the DDM as described in the previous section, a variable optical attenuator (VOA) reduces the number of photons per qubit down to the single photon level. The random pulse sequences for the two RF inputs of the DDM are provided by a 2.5~GHz transceiver driven by a hardware key distillation engine based on a field-programmable gate array (FPGA)(Xilinx Virtex 5), such that the separation $\tau$ between two time bins is 800~ps (1.25~GHz). The key distillation engine is seeded with a Quantum Random Number Generator (QRNG) (ID Quantique Quantis). The qubit frequency is 625~MHz for COW and BB84 and 1.25~GHz for DPS. The RF pulses are synchronized with an external variable RF delay and amplified by means of two 12.5~GHz RF modulator drivers (Picosecond 5865). The dual-drive integrated LiNbO$_{3}$ modulator, is the Photline MZDD-LN-10. It is biased with two DC voltages ($V_{\text{DC,1}}$ and $V_{\text{DC,2}}$) in order to have a phase difference of $\pi$ when the electrical pulses are off, leading to no light at the output.

\begin{figure}[tbp]
		\centering
		\includegraphics[width=0.8\textwidth]{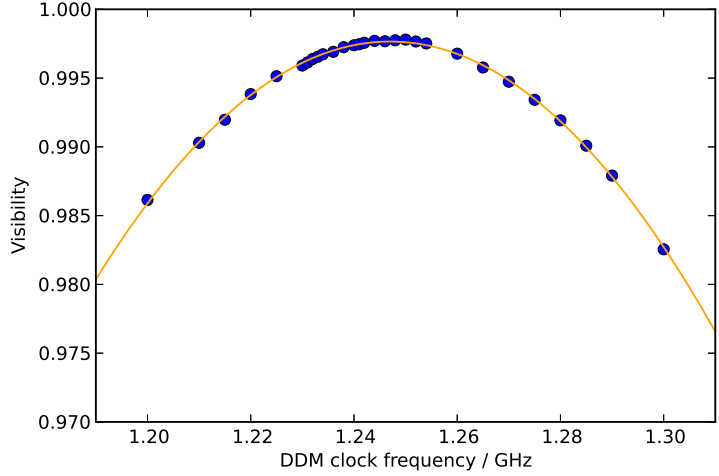}

\caption{Sweep of the DDM clock frequency to find the optimum intereferometer visibility. Correspondingly, the chosen operating frequency was 1.25~GHz, with a raw visibility of $99.76\pm0.04\%$.}
\label{fig:Inter_vis}
\end{figure}

As mentioned previously, an attractive feature of our transmitter is that it can be easily adapted to work with interferometers of different imbalances. Moreover one can take advantage of this fact to loosen the accuracy requirements in production of fiber interferometers. Indeed manufacturing a fiber interferometer with a precise imbalance is non-trivial as it requires accurate fiber cleaving with an error on the order of tens of microns. Instead one could manufacture an interferometer with limited accuracy, analyze its frequency response and then adapt the transmitter frequency to the optimum. By driving the DDM with a sine function, we measured the resulting visibility as a function of frequency.  Figure~\ref{fig:Inter_vis} shows that the maximum visibility for our interferometer is very near 1.25~GHz, hence this is the frequency we selected for our experiments. From this data we calculated that in order to achieve a QBER within 0.1\% of the interferometer optimum, an accuracy of 20~MHz in the clock frequency, which is easily achievable with our system. The raw visibility at 1.25~GHz is $99.76\pm0.04\%$ from which we would expect a phase error rate of only about $0.12\%$ when running QKD, which results from interferometer imperfections and a finite coherence length of the laser.   

%%%%%%%%%%%%%%%%%%%%%%%%%%%%% Fig. Interferometer Visibility %%%%%%%%%%%%%%%%%%%%%%%%%%%%%%%%%%%%%%%%%%%%
%============================================================================================

\begin{figure}[tbp]
	\begin{minipage}[b]{0.51\textwidth}
		\centering
		\includegraphics[width=1.1\textwidth]{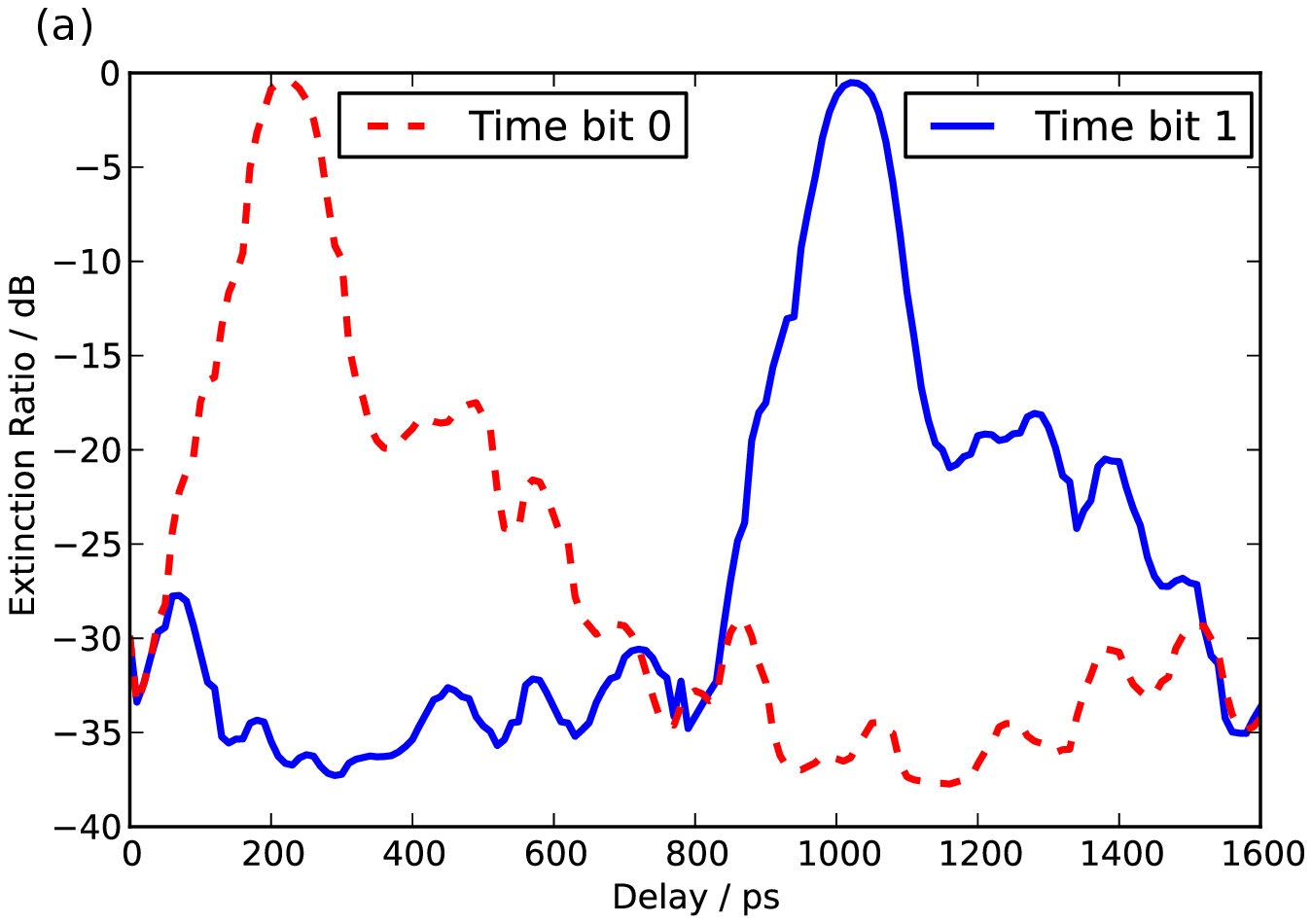}
	\end{minipage}
	\hspace{0.05cm}
	\begin{minipage}[b]{0.51\textwidth}
		\centering
		\includegraphics[width=1.1\textwidth]{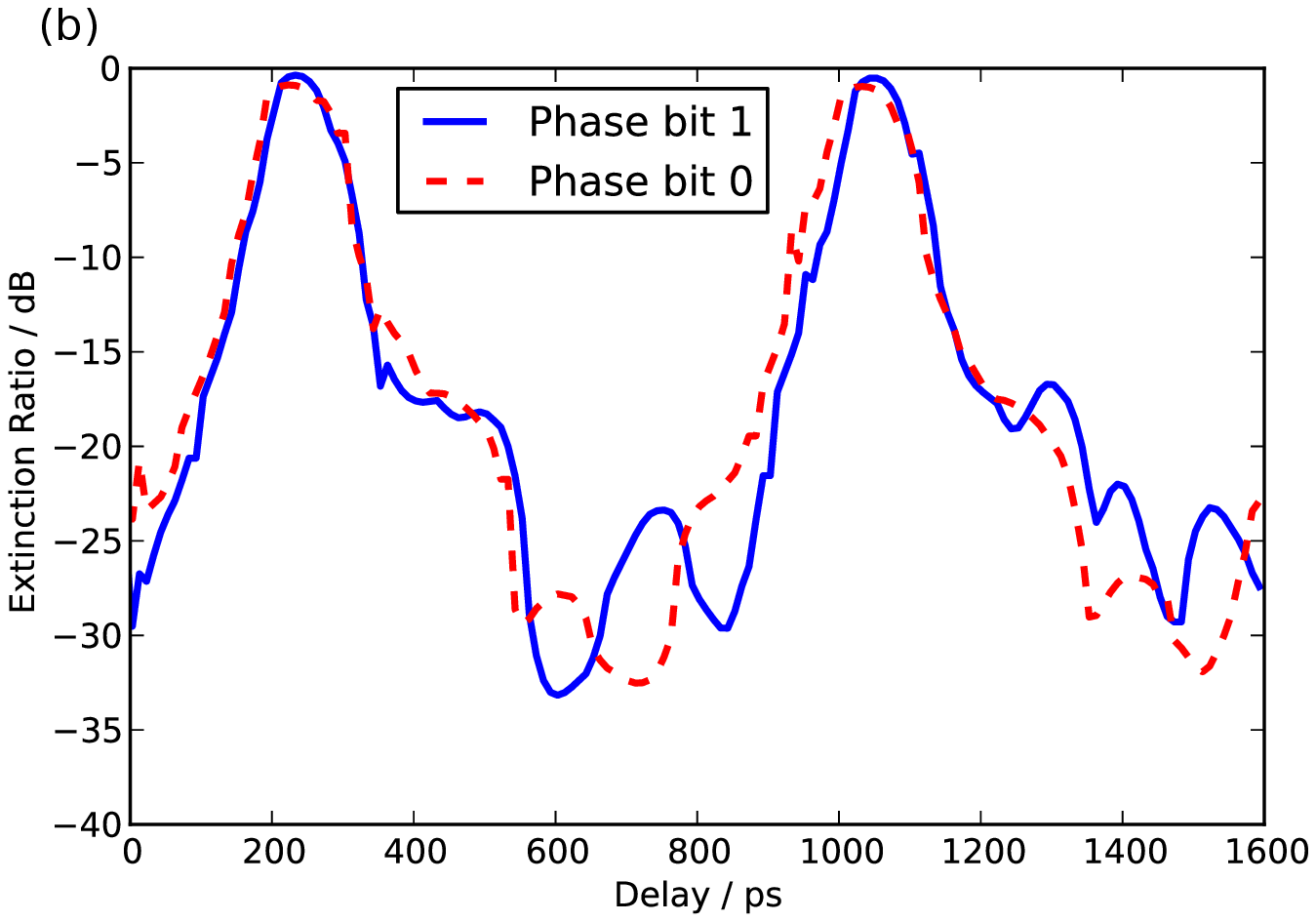}
	\end{minipage}

\caption{(a) Time basis extinction ratio for the two bit values encoded with light being sent in the early or late time bins, created using push-pull operation of the DDM as for COW and the BB84 time basis. (b) Phase basis pulse shape with bits encoded in the relative phase of 0 or $\pi$ between two pulses, created using RF drive pulses of maximum amplitude as used for the DPS protocol.}
\label{fig:Extinction}
\end{figure}

In view of QKD, it is important that the pulses generated by the transmitter are indistinguishable, i.e. the shape is independent of the bit value or the history of previous bit values. Mismatch in pulse shape for states encoded in the phase basis leads to errors due to imperfect interference. Moreover, a high extinction ratio is desirable, i.e. a good suppression of light in empty time bins, since this minimizes the error rate in the time basis. A measurement of the extinction ratio was carried out by scanning a 33~ps pulsed laser with respect to a fixed qubit pattern generated by the DDM and monitoring the output power as a function of delay. Figures~\ref{fig:Extinction}(a) and \ref{fig:Extinction}(b) show, on a logarithmic scale, pulse shapes of states encoded in the time and phase bases, which have a FWHM duration of approximately 90~ps. For phase encoding of $0$ and $\pi$ the pulse shapes remain very similar with significant differences only occurring below -20 dB. In the time basis, the extinction ratio is $>$27 dB which should result in a QBER due to imperfect amplitude modulation of less than 0.2\%.

Running the multi-protocol QKD platform we were able to test the performance of the DDM transmitter in a key exchange scenario. The receiver was configured accordingly to the protocol being tested, as illustrated in Fig.~\ref{fig:DDMnetwork}, with the phase basis ($\sigma_{\text{X}}$) measurement carried out using a fiber-optic Michelson interferometer with Faraday mirrors, rendering it polarization insensitive. To compensate for phase changes in the interferometer due to temperature fluctuations, it was possible to finely tune the laser current in order to control the optical wavelength, thus the phase error rate was stabilized effectively. Additionally, the errors were minimized by actively tracking the $V_{\text{DC,1}}$ of the DDM. For BB84 and COW, a passively random measurement basis choice was achieved using a 3~dB coupler before the interferometer. Single photon detection was carried out with free-running InGaAs APD detectors (ID Quantique ID220) \cite{lunghi2012a}, operated at 20\% detection efficiency with 20~$\mu s$ deadtime in order to reduce the afterpulsing probability below 1\%. At these settings the timing jitter of the detectors was around 250~ps (FWHM) and the observed dark count rate was in the region of 1.5~kHz. Detections on Bob's side were registered by the FPGA and the relevant sifting information exchanged over a 2.5~Gb/s classical optical communication link (Finisar FWLF). The fiber length between the source and receiver was only a few meters, however the variable optical attenuator was used to change the average number of photons per qubit at the receiver between 1 and $10^{-5}$, in order to simulate operation over various channel losses.

%%%%%%%%%%%%%%%%%%%%%%%%%%%%% Fig. Extinction Ratio %%%%%%%%%%%%%%%%%%%%%%%%%%%%%%%%%%%%%%%%%%%%
%============================================================================================

%%%%%%%%%%%%%%%%%%%%%%%%%%% Fig. QKD over channel losses %%%%%%%%%%%%%%%%%%%%%%%%%%%%%%%%%%%%
\begin{figure}[htbp]
	\centering
	\begin{minipage}[b]{0.7\textwidth}
		\includegraphics[width=\textwidth]{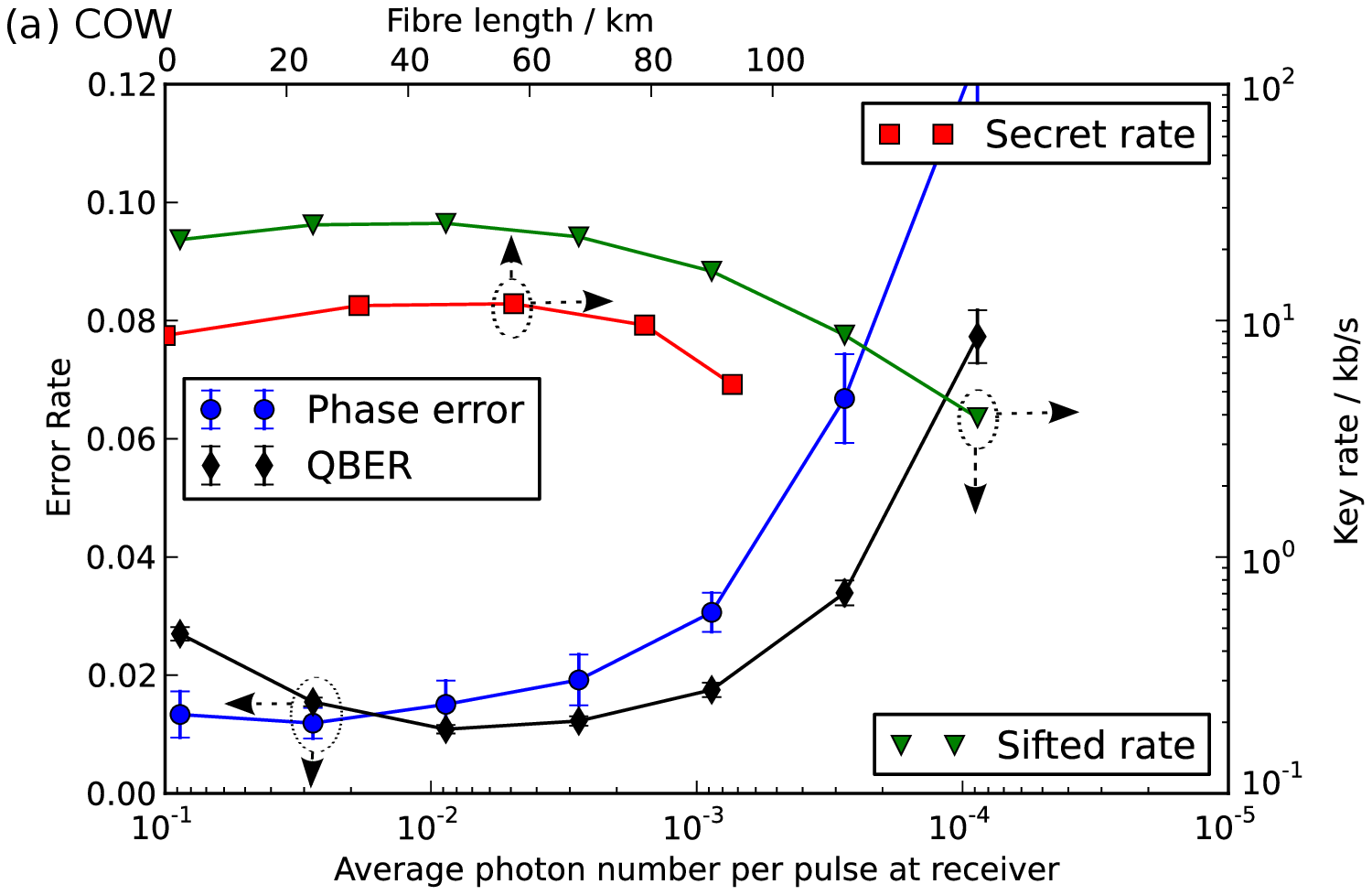}
	\end{minipage}
	\hspace{0cm}
		\centering
	\begin{minipage}[b]{0.7\textwidth}
		\includegraphics[width=\textwidth]{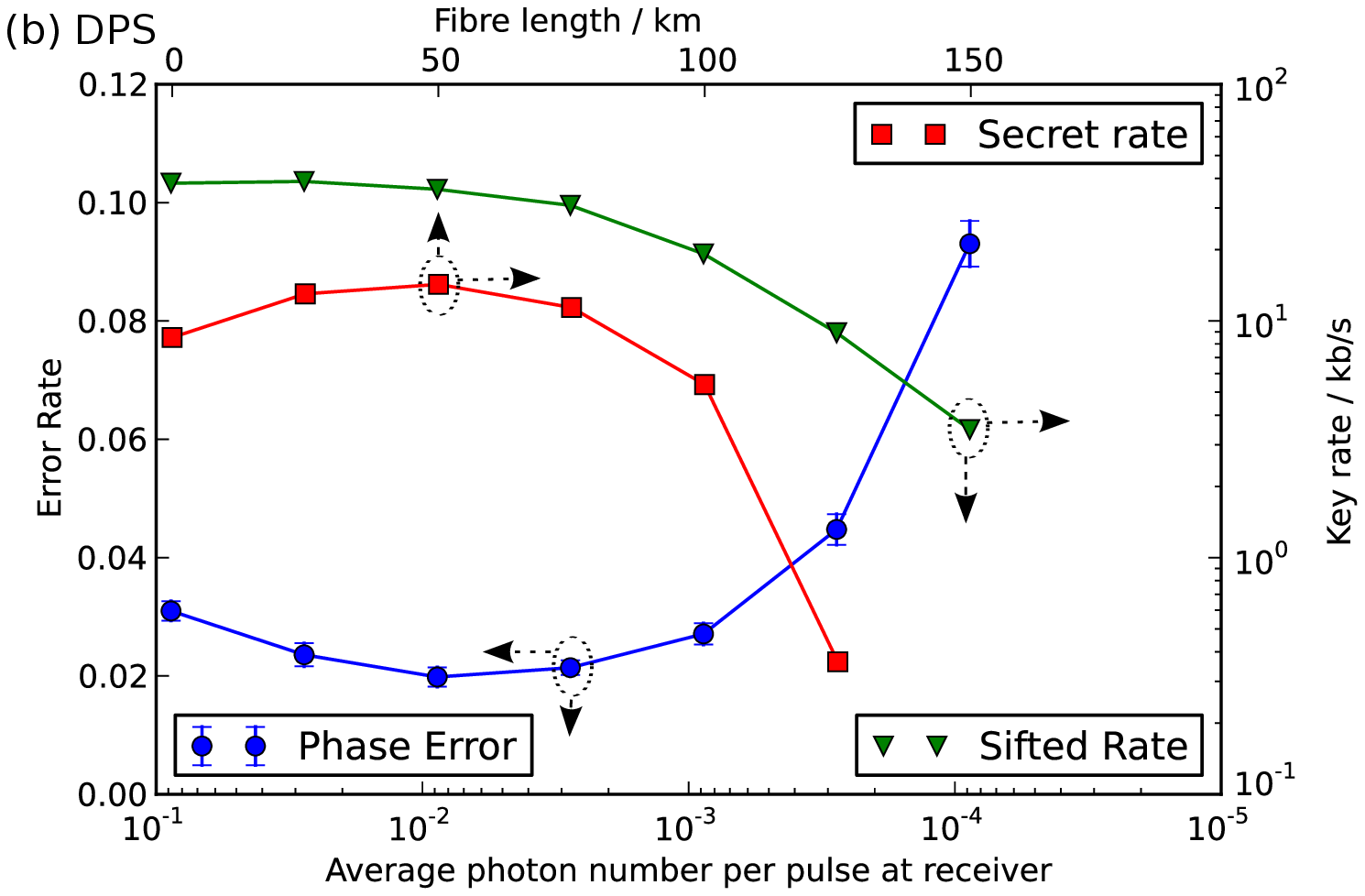}
	\end{minipage}
	\hspace{0cm}
	
	\centering
	\begin{minipage}[b]{0.7\textwidth}
		\includegraphics[width=\textwidth]{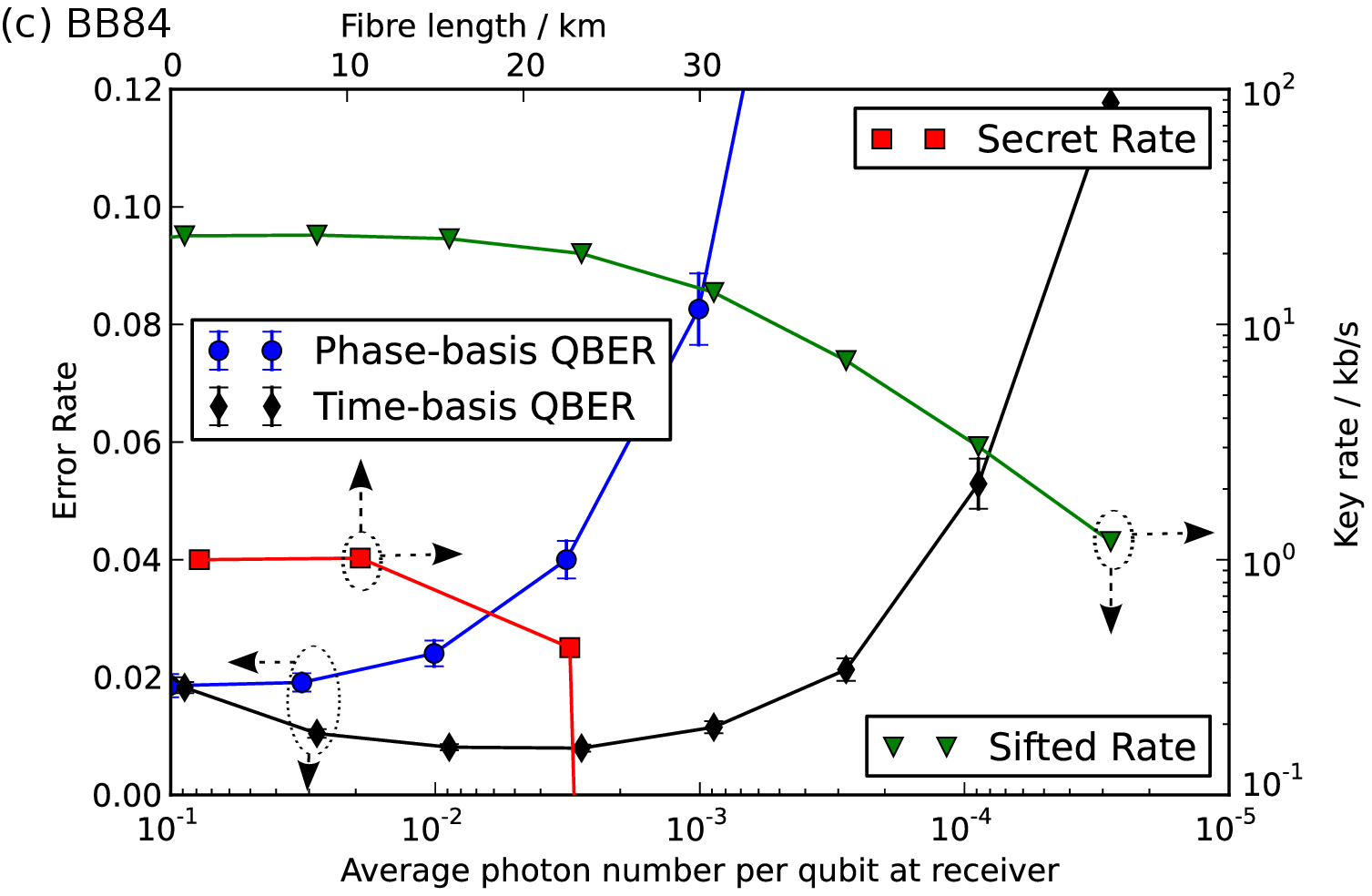}
	\end{minipage}
	
\caption{(a) Measured sifted rate, secret rate, QBER and visibility (converted into phase-error rate for direct comparison) using the COW protocol. (b) Sifted rate, secret rate and phase-error rate for the DPS protocol for different losses in the quantum channel. (c) Measured QBER in the time and phase bases along with the sifted and secret rates for randomly prepared pulse sequences using the BB84 protocol. The sifted rates in each figure are plotted as a function of the photon number at the receiver (bottom axis), whilst the secret key rates are presented as a function of fibre length (top axis) assuming losses of 0.2~dB/km.}
	\label{fig:QBER}
\end{figure}
%=============================================================================================

Using the BB84 protocol, the basis choice and bit value sent by Alice were chosen at random with equal probabilities. For the COW protocol, the transmitted bit value was also chosen randomly, whilst the decoy sequence was sent with a probability of 15.5\%. All of the detection events on Bob's side were directly compared to Alice's choice over a service channel of the key distillation engine in order to obtain the true value of the error rate. The DPS protocol was tested using predetermined bit sequences. Figures~\ref{fig:QBER}(a)-\ref{fig:QBER}(c) show the QBER and sifted rate for different losses in the quantum channel, whilst using the 3 different protocols. From this data, an estimated secret key rate that would be achievable in a full QKD scenario is calculated as a function of fibre length, according to the security proofs outlined in Section~\ref{sec:state_prep}. Please note that the distance scale on the top of Figs.~\ref{fig:QBER}(a)-\ref{fig:QBER}(c) is different for each protocol since the optimum photon number at the transmitter varies a lot according to the transmission distance and the protocol being used. These results show the full system performance, including detector dark counts. 

The minimum QBER value for each protocol drops below 1\% and 2\% for the time and phase basis, respectively.  For BB84, the phase error curve is displaced to higher photon numbers compared to that of the time basis. This is partially due to the additional losses in the interferometer and circulator along with the fact that only one time-bin per qubit is monitored corresponding to the instance where the early and late parts of the qubit take the long and short interferometer arms respectively, whilst any other combination is discarded. With decreasing photon numbers, the dark counts quickly became the dominant contributor to the QBER, whilst at high photon numbers the QBER also increased due to saturation effects in the detectors. This is the reason for the counter intuitive reduction of sifted and secret rates at high photon numbers. The maximum sifted rates saturated at just below 25~kb/s (COW and BB84) and 40~kb/s (DPS), due to the deadtime of the detectors. 

It is evident that the secret key rates for the three protocols are quite different. One must keep in mind that the presented secret rates and distances are only an indication of what could be achievable, calculated from the sifted rates and the QBER. COW was analyzed taking finite key effects into account, assuming a block size of $10^6$, whilst treatment of DPS and BB84 was carried out in the asymptotic limit. DPS performs the best in terms of distance, allowing transmission beyond 100~km, whilst COW reaches a distance of about 90~km. Due to the fact that BB84 with weak coherent states is susceptible to the photon-number splitting attack it exhibits the worst secret key rate. Additionally the nonrandom phase limits the distance to about 20~km. We expect that with phase randomization the distance would increase by around a factor of 3. In order to increase the distance further, in would be logical for future work to investigate the possibility of implementing the decoy state version of BB84 with this transmitter.

Although this work does not aim to be a review of the different protocols (we would direct the reader to \cite{Scarani09} for this), it is clear that they posses various advantages and disadvantages, be it achievable distance, secret key rate, ease of implementation or security against a given attack. This highlights the appeal of having a flexible transmitter that could allow adaptation to the most efficient protocol in a given scenario. In this investigation we did not implement all of the real-time post processing required for full QKD, however, in practice our transmitter is directly compatible with existing QKD platforms. Currently, switching protocols with our system must be done manually and takes a few tens of minutes in order to reconfigure the FPGA. Since the majority of the resource heavy processes such as error correction and privacy amplification are independent of the protocol, in should be possible to implement a system that could automatically switch protocols and receivers within tens of seconds. For real-world use, in order to increase the secret key rate and achievable distance dramatically, the main change that would need to be made to the overall system is the use of rapid-gated APD detectors \cite{walentaSine, restelli2013}, or superconducting nanowire single photon detectors \cite{marsili2013}. 

In scope of this investigation the figure of merit relating directly to the transmitter performance is the optical quantum bit error rate (${\text{QBER}_{\text{opt}}}$), which is calculated by subtracting the detector dark count contribution. By taking into account the detector deadtime, we calculate the ${\text{QBER}_{\text{opt}}}$ for each protocol in the range of $10^{-2} - 10^{-3}$ photons per pulse/qubit, where high detection rate was observed giving the best statistics, without detector saturation effects. The results are summarized in Table~\ref{table:OptQBER}.

%%%%%%%%%%%%%%%%%%%%%%%%%%%%%%% Table: Optical QBER %%%%%%%%%%%%%%%%%%%%%%%%%%%%%%%%%%%%%%%%%%%%

\begin{table}[htb]
\centering\caption{Optical QBER.}
\label{table:OptQBER}
\begin{tabular}{| l | p{2.1cm} | p{2.1cm} |}
\hline 
Protocol & Phase basis ${\text{QBER}_{\text{opt}}}$ & Time basis ${\text{QBER}_{\text{opt}}}$ \\ \hline \hline
DPS & 1.83~$\pm$~0.19\% & N/A \\ \hline
COW & 0.92~$\pm$~0.41\% & 0.89~$\pm$~0.08\% \\ \hline
BB84 & 1.51~$\pm$~0.16\% & 0.58~$\pm$~0.06\% \\ \hline 
\end{tabular}
\end{table}

%===============================================================================================

As expected from the extinction ratio measurement, the resulting time basis error rate is very low; a portion of which can still be attributed to the timing jitter and afterpulsing of the detectors. The main contribution to the phase error rate is expected to stem from pulse-to-pulse amplitude and shape fluctuations (signal eye spreading) which lead to imperfect interference at the receiver. The fact that the phase error rate for COW is lower than that of BB84 and DPS means that more errors arise from the state with a phase difference of $\pi$, since this is the only one not used in COW. This can be expected since this state is generated by first applying a phase shift in one arm of the DDM and then the other. Any mismatch in the electronic pulses driving each arm will directly result in interference errors. It is envisaged that further improvements to the pulse generator cards could reduce this value, by reducing the electronic pulse ripple, which leads to the significant distortions on the falling edge of the optical pulses, as seen in Fig.~\ref{fig:Extinction}(b). 

\begin{figure}[tbp]
	\begin{minipage}[b]{0.49\textwidth}
		\centering
\includegraphics[width=1.05\textwidth]{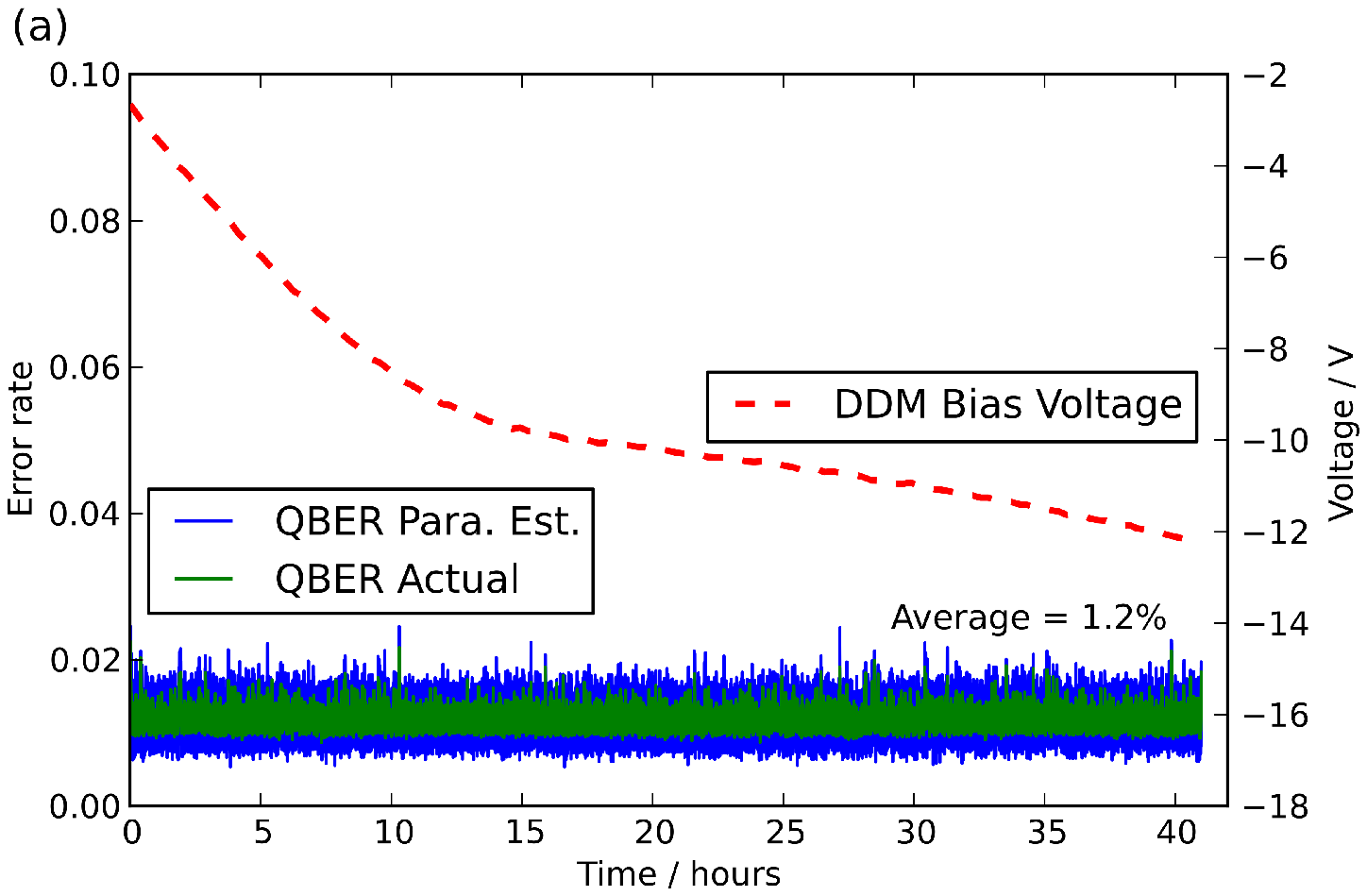}
	\end{minipage}
\hspace{0.25cm}
	\begin{minipage}[b]{0.49\textwidth}
		\centering
		\includegraphics[width=1.05\textwidth]{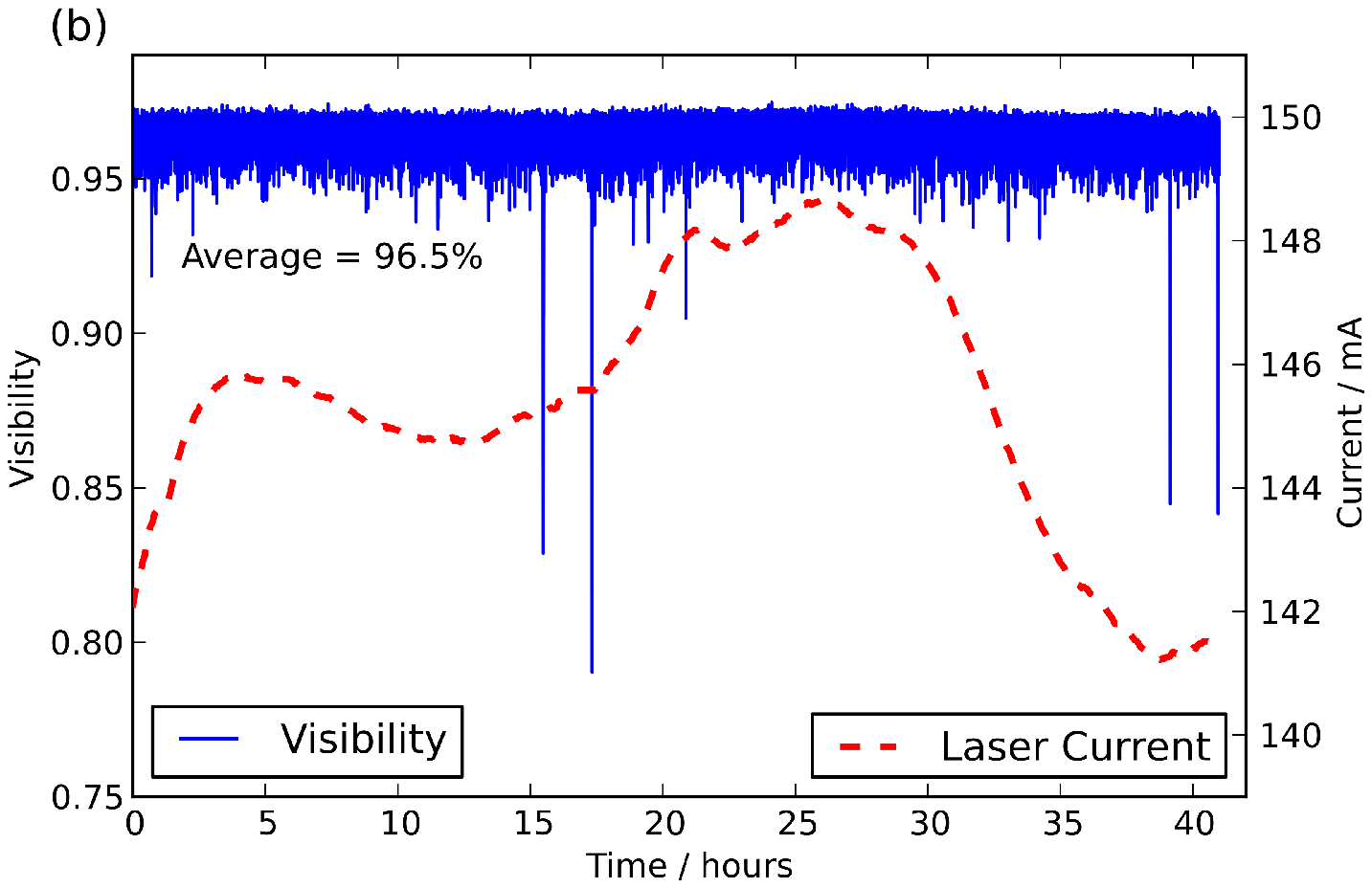}
	\end{minipage}
	\caption{(a) QBER measurement for COW protocol over a period of 40 hours, showing a stable average value of 1.2\% obtained with parameter estimation on 12.5\% of the sifted bits. This value is also verified separately by utilising the full detection statistics (QBER Actual). DDM bias voltage tracking is also shown, which was the main variable parameter used for QBER minimization. (b) Stable visibility measurement over the same 40 hour period, with an average value of 96.5\%. In this case the laser current was adjusted for stabilization. }
	\label{fig:QBER_long}
\end{figure}

In order to demonstrate the stable operation of the transmitter, a constant photon number per pulse of $10^{-2}$ was set, whilst using the COW protocol. Active tracking of the DDM bias voltage and the laser current was also enabled. During this measurement, parameter estimation was carried out over the authenticated public channel, on 12.5\% of the sifted bits, along with full QBER verification over the service channel of the key distillation engine. Figures~\ref{fig:QBER_long}(a) and \ref{fig:QBER_long}(b) show operation over a period of 40 hours with an average QBER of 1.2\% and a visibility of 96.5\%. The resulting bias voltage and laser current is also plotted. This demonstrates that our system is very robust and can easily track and optimize the required parameters. 

%%%%%%%%%%%%%%%%%%%%%%%%%%%%%%%%%%%%%%%%% Fig. Stability %%%%%%%%%%%%%%%%%%%%%%%%%%%%%%%%%%%%%%%%
%=================================================================================================

%=================================================================================================
%%%%%%%%%%%%%%%%%%%%%%%%%%%%%%%%%%%%%% Conclusion %%%%%%%%%%%%%%%%%%%%%%%%%%%%%%%%%%%%%%%%%%%%%%%%
%=================================================================================================

\section{Conclusion}
We have demonstrated a novel optical quantum communication transmitter that can be flexible in terms of the QKD protocol that it uses. Moreover, the transmitter is constructed with off-the-shelf components, i.e. a commercial DDM running at a frequency of 1.25~GHz, a CW laser and a VOA. The experimental results show an extinction ratio of $>27$ dB allowing time and phase coding with low QBER for three different protocols, namely BB84, COW and DPS. The performance of the transmitter was analyzed with the three protocols by running QKD over different channel losses. Indicative secret key rate and achievable distances were presented. In future work it would be possible to automatically switch between different receivers dedicated to different protocols, when using a reconfigurable network. Switching between receivers with different interferometer path differences could also be possible. The source can simplify extensions to network QKD environments providing a reduction in system complexity, management and cost. 

%=================================================================================================
%%%%%%%%%%%%%%%%%%%%%%%%%%%%%%%%%%%%%% Acknowledgements %%%%%%%%%%%%%%%%%%%%%%%%%%%%%%%%%%%%%%%%%%%
%=================================================================================================

\section*{Acknowledgments}
We wish to thank Charles Ci Wen Lim for helpful advice and Antonio Ruiz-Alba for stimulating discussions and involvement with our group leading to this investigation. This work was supported by the Swiss National Centre of Competence in Research, Quantum Science and Technology project (NCCR-QSIT).

\end{document}